\documentstyle[epsfig,aps,prb,twocolumn,floats,eqsecnum]{revtex}

\begin{document}

\newcommand{\gsim}{
\,\raisebox{0.35ex}{$>$}
\hspace{-1.7ex}\raisebox{-0.65ex}{$\sim$}\,
}

\newcommand{\lsim}{
\,\raisebox{0.35ex}{$<$}
\hspace{-1.7ex}\raisebox{-0.65ex}{$\sim$}\,
}

\newcommand{\const}{ {\rm const} }
\newcommand{\arctanh}{ {\rm arctanh} }

\newcommand{\onehalf}{\mbox{\scriptsize 
\raisebox{1.5mm}{1}\hspace{-2.7mm}
\raisebox{0.5mm}{$-$}\hspace{-2.8mm}
\raisebox{-0.9mm}{2}\hspace{-0.7mm}
\normalsize }}

\newcommand{\fivehalf}{\mbox{\scriptsize 
\raisebox{1.5mm}{5}\hspace{-2.7mm}
\raisebox{0.5mm}{$-$}\hspace{-2.8mm}
\raisebox{-0.9mm}{2}\hspace{-0.7mm}
\normalsize }}

\newcommand{\onethird}{\mbox{\scriptsize 
\raisebox{1.5mm}{1}\hspace{-2.7mm}
\raisebox{0.5mm}{$-$}\hspace{-2.8mm}
\raisebox{-0.9mm}{3}\hspace{-0.7mm}
\normalsize }}

\newcommand{\twothird}{\mbox{\scriptsize 
\raisebox{1.5mm}{2}\hspace{-2.7mm}
\raisebox{0.5mm}{$-$}\hspace{-2.8mm}
\raisebox{-0.9mm}{3}\hspace{-0.7mm}
\normalsize }}

\newcommand{\fourthird}{\mbox{\scriptsize 
\raisebox{1.5mm}{4}\hspace{-2.7mm}
\raisebox{0.5mm}{$-$}\hspace{-2.8mm}
\raisebox{-0.9mm}{3}\hspace{-0.7mm}
\normalsize }}

\newcommand{\onefourth}{\mbox{\scriptsize 
\raisebox{1.5mm}{1}\hspace{-2.7mm}
\raisebox{0.5mm}{$-$}\hspace{-2.8mm}
\raisebox{-0.9mm}{4}\hspace{-0.7mm}
\normalsize }}

\newcommand{\oneeights}{\mbox{\scriptsize 
\raisebox{1.5mm}{1}\hspace{-2.7mm}
\raisebox{0.5mm}{$-$}\hspace{-2.8mm}
\raisebox{-0.9mm}{8}\hspace{-0.7mm}
\normalsize }}

\bibliographystyle{prsty}

\title{ \begin{flushleft}
{\small \em submitted to}\\
{\small 
PHYSICAL REVIEW B 
\hfill
VOLUME  XXX
NUMBER XXX
\hfill 
MONTH XXX
}\\
\end{flushleft}  
Curie temperature of anisotropic ferromagnetic films
}

\author{
D.~A. Garanin \cite{e-gar}
}

\address{
Max-Planck-Institut f\"ur Physik komplexer Systeme, N\"othnitzer Strasse 38,
D-01187 Dresden, Germany\\
\smallskip
{\rm (Received 16 June 1998) }
\bigskip\\
\parbox{14.2cm}
{\rm
Dimensional crossover of ordering in ferromagnetic films with both periodic
and free boundary conditions is studied for the exactly
solvable uniaxial model of classical $D$-component spin vectors in the limit
$D\to\infty$.
Analytical and numerical solution of the exact equations describing this model
shows that for lattice dimensionalities $d>4$, finite-size corrections  to the bulk values
of $T_c$ are characterized by the  mean-field exponents and
anisotropy-dependent amplitudes.
For $d\leq 4$, the mean-field behavior is only realized in the 
region $\kappa_c N \gg 1$, where $\kappa_c$ is the dimensionlesss
inverse {\em transverse} (with respect to the easy axis) bulk correlation
length at $T_c$ and $N$ is the number of layers in the film.
In the region $\kappa_c N \ll 1$ and the dimensionality range $3 < d \leq 4$,
finite-size corrections are described by the universality class of the
isotropic $D=\infty$ model. 
For $d\leq 3$, magnetic ordering vanishes in the isotropic limit,
$\kappa_c \to 0$, since the film behaves as an object of dimensionality
$d'=d-1 \leq 2$ and long-wavelength fluctuations destroy the order.
Here the suppression of $T_c$ of a film can be substantial, depending on the
competition between the weakening anisotropy and the increasing film
thickness. 
For thick films, $T_c$ becomes small only for very small anisotropy.
\smallskip
\begin{flushleft}
PACS number(s): 64.60.Cn, 75.10.Hk, 75.30.Pd, 75.70.Ak 
\end{flushleft}
} 
} 
\maketitle

\section{Introduction}

Ferromagnetic ordering in the film geometry shows a dimensional crossover
on the film thickness, which is measured by the number of layers $N$ for the
lattice spacing $a_0$ set to unity, and on the closeness
to the Curie point $T_c$.
For a spatial dimensionality $d$, thick films, $N\gg 1$, show a
$d$-dimensional behavior  far enough from $T_c$.
On the other hand, in the close vicinity of $T_c$ the film behaves as an
object characterized by the reduced dimensionality $d'=d-1$.
The latter regime is realized for $\xi_c \gsim N$, where $\xi_c$ is the
dimensionless bulk correlation length.  
An important implication of this dimensional crossover is that isotropic
Hamiltonians cannot describe ordering of films in three dimensions.
Indeed, since films behave two dimensionally at the would be critical point,
the long-wavelength fluctuations (spin waves, Goldstone modes) preclude ordering.
Due to the exponential increase of the correlation length with lowering
temperature, two-dimensional magnets are extremely sensitive to the uniaxial
anisotropy which stabilizes order.
It is intuitively clear that the stabilizing effect of the anisotropy in thick films
should be much stronger than that in usual two-dimensional systems.

Finite-size effects and crossovers between different universality classes in thin magnetic films 
have been observed in a number of experiments.
\cite{neu62,gra74,dueetal89,libab92,schsierie96,bab97}
The suppression of $T_c$ and other effects in films of arbitrary thickness $N$
are frequently addressed with the spatially-inhomogeneous mean-field approach
of Ref. \onlinecite{woldewhalpal71}, which, naturally, misses the influence of
Goldstone modes on ordering mentioned above.
The general large-$N$ asymptotic form of the $T_c$-shift in thick films is given by
%
\begin{equation}\label{TcShNbig}
[T_c(\infty) - T_c(N)]/T_c(\infty) \cong A/N^\lambda, 
\end{equation}
as was initially established by the high-temperature series expansions (HTSE)
for the Ising model. \cite{all70,fis71}
For the exponent $\lambda$ the finite-size scaling theory
\cite{fisbar72,bar83ptcp} yields
$\lambda=1/\nu_b$, where  $\nu_b$ is critical index for the bulk correlation length.
The coefficient $A$ in Eq. (\ref{TcShNbig}) should depend on the anisotropy.
For spatial dimensions $d\le 3$, it should strongly increase with approaching
the isotropic limit, so that for any thickness $N$ the right-hand side (rhs) of
Eq. (\ref{TcShNbig}) eventually approaches the value of 1.
In this case of a substantial suppression of $T_c$, the functional form of
Eq. (\ref{TcShNbig}) is no longer valid.
For very small uniaxial anisotropy, the film orders at such low temperatures that all
spins along the direction perpendicular to the surface are strongly correlated
and can be considered as a single composite spin.
Thus the film is mapped on $d'$-dimensional monolayer with the exchange
interaction $NJ$.
For the model with a uniaxially anisotropic exchange, the anisotropy of the exchange
interaction $\eta' \ll 1$  is replaced by $d\eta'/(d-1)$, since the composite spins acquire an
``internal'' anisotropy.
In this case one has
%
\begin{equation}\label{TcAnlo}
T_c[d,J,\eta',N] \cong T_c[d-1,NJ,d\eta'/(d-1),1].
\end{equation}
For the model with a single-site anisotropy, the latter, relative to the
exchange, is not renormalized.
It seems that no one of conventional theories can reproduce the crossover
between the two regimes described by Eqs. (\ref{TcShNbig}) and (\ref{TcAnlo}).

The effects of would be Goldstone modes on ordering in weakly-anisotropic
low-dimensional magnetic structures can be conveniently studied within the
exactly solvable anisotropic spherical model (ASM) which is the limit
$D\to\infty$ of the classical $D$-component spin-vector model.
The Hamiltonian of the latter has the form
%
\begin{equation}\label{dham}
{\cal H} = 
- \frac{1}{2}\sum_{ij}J_{ij}
\left(
m_{zi}m_{zj}  
+ \eta \sum_{\alpha=2}^D m_{\alpha i} m_{\alpha j}
\right) ,
\end{equation}
where ${\bf m}_i$ is the normalized $D$-component vector, 
$|{\bf m}_i|=1$, and $\eta \equiv 1-\eta' \leq 1$ is the dimensionless anisotropy factor. 
In the isotropic case, $\eta=1$, the partition function of this model
coincides in the limit $D\to\infty$ (Ref. \onlinecite{sta68pr}) with that of
the spherical model (SM).  \cite{berkac52} 
There are, however, a number of essential differences between both models.
In particular, the spherical model yields unphysical negative coefficient $A$ in
Eq. (\ref{TcShNbig}) for the film with free boundary conditions (fbc) because
of the global spin constraint. \cite{barfis73}
Improved versions of the SM without the global spin constraint
\cite{kno73,cosmazmih76} yield positive $A$, as it should.
However, within the SM the problem can be only considered in four dimensions,
since the SM cannot be done anisotropic and in three dimensions $T_c$ is zero
for all finite values of $N$.

By contrast, the ASM describes ordering of bulk samples in low
dimensions, \cite{garlut84d} since anisotropy gives rise to the gap for spin fluctuations.
In Ref. \onlinecite{gar97zpb} it was shown that even in the
isotropic case there are different, longitudinal and transverse correlation
functions (CF) in the ASM below $T_c$, in contrast to the inherently single CF
in the SM.
The ASM correlation functions have a kind of spin-wave form, which is,
however, valid in the whole range of temperatures.
This makes the ASM a convenient tool for the investigation of classical spin
systems at elevated temperatures at the level beyond the mean-field
approximation (MFA).
Whereas the critical coupling of fluctuations dies out in the limit
$D\to\infty$ and the critical behavior is simplified, the not less important
{\em qualitative} effects of the would be Goldstone modes on ordering are accounted for in the
``pure'' form.
The most recent example of the application of the ASM to the
spatially-homogeneous systems 
is Ref. \onlinecite{garcan98prb}, where the spin-liquid state of the classical
antiferromagnet on the kagom\'e lattice is considered.

In the inhomogeneous case, the closed lattice-based system of equations
describing the ASM was obtained in Ref. \onlinecite{gar96jpa}.
This system of equations has been analytically solved in the domain-wall
geometry for the {\em biaxial} generalization of the Hamiltonian (\ref{dham}).
This model features a phase transition of an interface, which can be described 
analytically beyond the MFA and observed in experiment. \cite{koegarharjah93}
A more complicated situation is realized in the semi-infinite geometry, where
analytical solutions can be only obtained in particular and limiting cases and
application of  numerical methods is necessary.
In Ref. \onlinecite{gar98pre} the correlation functions of the semi-infinite
ferromagnet have been computed at $T\geq T_c$ in the case of the so-called
ordinary phase transition (the transition at the surface, that is characterized
by its own set of critical indices, is driven by the phase transition in the bulk).
For the film geometry, the problem becomes even more difficult because of the
additional effect --- suppression of $T_c$ in comparison to its bulk value.
In my preceding Letter on ferromagnetic films, Ref. \onlinecite{gar96jpal}, only some analytical
results for $N\gg 1$ have been obtained, but no numerical calculations have
been performed.

The aim of this paper is a complete solution for the correlation functions and
Curie temperatures of ferromagnetic films within the ASM in the whole range of
parameters with the use of the numerical algorithm of  Ref. \onlinecite{gar98pre}.
In contrast to Ref. \onlinecite{gar96jpal}, calculations will be performed for
the model with a continuous spatial dimensionality $d$, which has been
introduced in Ref. \onlinecite{gar98pre}.
The latter is interesting, in particular, since in the isotropic limit nontrivial finite-size
corrections to $T_c$, which differ from the mean-field ones, are only realized
in the dimensionality range between three and four.
In addition, analytical resultes will be given for ferromagnetic
films consisting of a few layers.

The main part of this paper is organized as follows.
In Sec. \ref{basic} the basic equations describing the anisotropic spherical
model and the methods of their solution are be briefly reviewed.
In Sec. \ref{pbc} the analytical solutions for the simplified model with
periodic boundary conditions (pbc) are given.
In Sec. \ref{fbc} the available analytical solutions for the film with free
boundary conditions are considered.
In Sec. \ref{numres} the corresponding numerical results  are presented.
In Sec. \ref{discussion} the results obtained and some possible extensions of the
model are discussed.

\section{Basic relations}
\label{basic}

At or above $T_c$ in zero field, the magnetization  ${\bf m}_i$ is zero and the ASM is
described by the closed system of equations for the correlation functions of transverse
($\alpha\geq 2$) spin components, $s_{ij}\equiv D\langle m_{\alpha i}m_{\alpha j}\rangle$, 
and the spatially varying gap parameter, $G_i$.
More general forms of these equations with  ${\bf m}_i \ne 0$ and a lot of
other details can be found in Refs. \onlinecite{gar96jpa,gar96jpal} and
\onlinecite{gar98pre}; here only the most important formulas will be given.
In the geometry considered,  it is 
convenient to use the Fourier representation in $d'=d-1$ translationally invariant dimensions parallel to the surface and the site representation in the $d$th dimension. 
For the model  with nearest-neighbor (nn) interactions,
the equation for the Fourier-transformed CF $\sigma_{nn'}({\bf q})$ 
then takes the form of a system of the second-order finite-difference
equations in the set of layers $n=1,\,2,...,\infty$,
%
\begin{equation}\label{cffd}
2b_n\sigma_{nn'} - \sigma_{n+1,n'} - \sigma_{n-1,n'}  =
(2d\theta/\eta)\delta_{nn'} ,
\end{equation}
where $b_n$ is given by
%
\begin{equation}\label{bn}
b_n = 1 + d[(\eta G_n)^{-1}-1] + d'(1-\lambda'_{\bf q}) 
\end{equation}
and $\lambda'_{\bf q}$ for the $d$-dimensional hypercubic lattice reads 
%
\begin{equation}\label{lampr}
\lambda'_{\bf q} = \frac{1}{d'}\sum_{i=1}^{d'} \cos(q_i) .
\end{equation}
In Eq. (\ref{cffd}), $\theta$ is the reduced temperature defined by
%
\begin{equation}\label{thetadef}
\theta \equiv \frac{ T }{ T_c^{\rm MFA}(\infty) } , 
\qquad T_c^{\rm MFA}(\infty) = \frac{ 2dJ }{ D }
\end{equation}
The quantities $\sigma_{0,n'}$ and $\sigma_{N+1,n'}$ in the nonexisting
layers, which enter equations
(\ref{cffd}) at the film boundaries $n=1$ and $n=N$, are set to
%
\begin{equation}\label{bcond}
\begin{array}{ll}
\sigma_{0,n'} = \sigma_{N+1,n'} = 0, \qquad     & {\rm (fbc)} \\
\sigma_{0,n'} = \sigma_{N,n'}, \qquad \sigma_{N+1,n'} = \sigma_{1,n'}, 
\qquad & {\rm (pbc)}  
\end{array}
\end{equation}
as the free and periodic boundary conditions.
The autocorrelation functions in each of the $N$ layers, $s_{nn}$, satisfy the
set of constraint equations
%
\begin{equation}\label{sconstrfd}
 s_{nn} \equiv \int\!\!\!\frac{d^{d'}{\bf q}}{(2\pi)^{d'}} 
\sigma_{nn}({\bf q}) = 1 , 
\end{equation}
which are the consequence of the spin rigidity, $|{\bf m}_i|=1$.
A straightforward algorithm for the numerical solving the equations above
is, for a given set of $G_n$, to compute all $\sigma_{nn}$ from the system of
linear equations (\ref{cffd}) and then to put the results in
Eq. (\ref{sconstrfd}) to obtain, after the integration over the Brillouin zone, the set of nonlinear
equations for $G_n$.
After $G_n$ have been found, one can compute the longitudinal CF
$\sigma_{nn'}^{zz}$ from Eqs. (\ref{cffd}) and  (\ref{bn}), where $\eta$ is set
to 1.

For thick films far from the boundaries, it is convenient to consider the
deviation form the bulk value
%
\begin{equation}\label{g1ndef}
G_{1n} \equiv G_n-G \ll 1,
\end{equation}
that is proportional [and for $T\leq T_c(\infty)$ equal] to the 
inhomogeneous part of the reduced energy pro site $\tilde U_{1n}$ (see Ref. \onlinecite{gar98pre}).
The bulk gap parameter $G$ satisfies the equation
%
\begin{equation}\label{GbulkEq}
\theta G P(\eta G) = 1,
\end{equation}
where 
%
\begin{equation}\label{px}
P(X) \equiv \int\!\!\!\frac{d^d{\bf k}}{(2\pi)^d}\frac{1}{1-X\lambda_{\bf k}} 
\end{equation}
is the lattice Green function.
The quantity $\lambda_{\bf k} \equiv J_{\bf k}/(2dJ)$ for the nn
interaction  is given by Eq. (\ref{lampr}) with $d'\Rightarrow d$ and ${\bf q} \Rightarrow \bf k$.
The solution $G$ of Eq. (\ref{GbulkEq}) increases with lowering temperature
$\theta$; at $G=1$ the gap in the longitudinal CF closes, longitudinal
susceptibility diverges, and the phase transition occurs.
This defines the bulk transition temperature  \cite{garlut84d}
%
\begin{equation}\label{thetacbulk}
\theta_c(\infty) = 1/P(\eta),
\end{equation}
that generalizes the well known result for the spherical model 
$\theta_c=1/P(1)$ (Ref.  \onlinecite{berkac52}).
The lattice Green function $P(X)$ satisfies $P(0)=1$ and has a singularity at $X\to 1$, the form of
which in different dimensions can be found in Ref. \onlinecite{gar98pre}.
For $d\leq 2$, the Watson integral $W\equiv P(1)$ goes to infinity; thus
formula (\ref{thetacbulk}) yields nonzero values of the Curie temperature only
for the anisotropic model, $\eta<1$.
It should be noted that in the anisotropic case the critical indices of the
model coincide with the mean-field ones due to the suppression of the
singularity of $P(\eta G)$ for $G\to 1$.
Below $\theta_c$, the spontaneous magnetization appears, and $G$ sticks to 1.

For $n\gg 1$, $q\ll 1$, and $\kappa\ll 1$, where $\kappa$ is the inverse
transverse correlation length defined by
%
\begin{equation}\label{kapdef}
\kappa^2 \equiv 2d[1/(\eta G) -1] \cong 2d[1-\eta G] \ll 1 ,
\end{equation}
one can reduce the second-order finite-difference equation (\ref{cffd}) to the differential equation for the Green's function 
%
\begin{equation}\label{cfdiffeq}
\left( \frac{d^2}{dn^2} - \tilde q^2 + 2dG_{1n} \right) \sigma_{nn'} = - 2d\theta\delta(n-n') ,
\end{equation}
where $n$ is considered as a continuous variable, $\tilde q \equiv\sqrt{\kappa^2+q^2}$,
and  $G_{1n}$ is defined by Eq. (\ref{g1ndef}). 
Note  that the transverse bulk correlation length 
$\xi_{c\perp} \equiv \xi_{c\alpha}\equiv 1/\kappa$ increases 
without diverging with decreasing temperature down to $\theta_c$ and 
remains constant below 
$\theta_c$, in accordance with the behavior of $G$.
The critical-point value 
%
\begin{equation}\label{kapcdef}
\kappa_c^2 \equiv 2d[1/\eta -1]
\end{equation}
 measures the anisotropy and
varies between 0 for the isotropic model and $\infty$ for the classical Ising model.
Analytical solution of Eq. (\ref{cfdiffeq}) is only possible in particular
cases, when $G_{1n}$ has a simple form and can be guessed.
In the domain-wall geometry, $G_{1n}\propto 1/\cosh^2[(n-n_0)/\delta]$, where
$n_0$ corresponds to the center of the domain wall and $\delta$ is the
self-consistently determined domain-wall width. \cite{gar96jpa}
Accordingly, the solution for $\sigma_{nn'}$  is a combination of hyperbolic
and exponential functions.
For the isotropic semi-infinite ferromagnet at the ordinary phase transition
in the dimensionality range $2<d<4$ the form of $G_{1n}$ is 
\cite{bramoo77prljpa,gar98pre}
%
\begin{equation}\label{g1nd24}
G_{1n} = \frac{ \onefourth-\mu^2 }{ 2dn^2 } , \qquad \mu = \frac{ d-3 }{ 2 } ,
\end{equation}
and $\sigma_{nn'}$ can be expressed through the modified Bessel functions
$I_\mu(qn)$ and $K_\mu(qn)$.

Note that long-wavelength equation (\ref{cfdiffeq}) is valid for continuous
lattice dimensionalities $d$ and its solution in the asymptotic region $n\gg 1$
is universal.
The disadvantage of Eq. (\ref{cfdiffeq}), however, is that it cannot be solved
numerically in confined geometries, which is desirable in most cases (e.g., for a semi-infinite
ferromagnet away from criticality) when there is no analytical solution.
The problem is the strong divergence of $G_{1n}$ near the boundaries [see Eq. (\ref{g1nd24})]
and the ensuing loss of the boundary condition.
In fact, however, this divergence is an artifact of the continuous
approximation.
The lattice-based calculation of Ref. \onlinecite{gar98pre}  for the simple cubic
lattice at isotropic criticality yields, on the
contrary, rather small values of $G_{1n}$ near the boundary: $G_{11}=0.09615$,
$G_{12}=0.01254$, $G_{13}=0.00532$, $G_{14}=0.00291$, etc.
To get access to continuous dimensionalities while perserving the discrete structure
of the lattice in the direction perpendicular to the surface, a special kind
of a ``lattice'' has been introduced in Ref. \onlinecite{gar98pre}.
This lattice is characterized by the Fourier-transformed exchange interaction
 $J_{\bf k} \equiv J_0 \lambda_{\bf k} $ of the form
%
\begin{equation}\label{lamcomb}
\lambda_{\bf k} \equiv \frac 1d \cos k_z + \frac{d'}{d}\lambda'_{\bf q},
\qquad  \lambda'_{\bf q} \Rightarrow 1-\frac{q^2}{2d'}
\end{equation}
[cf. Eq. (\ref{lampr})] with a continuous $d'=d-1$ and  {\em in the whole Brillouin zone}.
The natural hypercubic cutoff $|k_i|\leq \pi$ and the corresponding density of
states are modified for the ${\bf q}$ components according to
%
\begin{equation}\label{contint}
\int\!\!\!\frac{d^{d'}{\bf q}}{(2\pi)^{d'}} \ldots \Rightarrow 
\frac{d'}{\Lambda^{d'}}\int_0^\Lambda dq q^{d'-1}\ldots
\end{equation}
with $\Lambda = \sqrt{2(d+1)}$.
This choice of $\Lambda$ preserves the usual sum rules for the exchange interaction. 
Numerical calculations of Ref. \onlinecite{gar98pre} on the
continuous-dimension lattice confirmed formula (\ref{g1nd24}) in the
asymptotic region, $n\gg 1$.
Near the boundary, deviations from universality have been detected.
In particular, for the continuous dimensionality $d=3.0$ one obtains  $G_{11}=0.10672$,
$G_{12}=0.01195$, $G_{13}=0.00495$, $G_{14}=0.00270$, etc., which deviates
from the results for the simple cubic lattice quoted above.

If the gap parameter $G_n$ is uniform, as in the bulk or in a film with
periodic boundary conditions, Eq. (\ref{cffd}) can be solved analytically in a
straightforward way.
In other cases, an efficient numerical algorithm is needed.
One possibility, which is used in Ref. \onlinecite{gar98pre} and in the present article, is to represent the layer-layer autocorrelation function
$\sigma_{nn}$ in the continued-fraction form \cite{gar96jpal,gar98pre}
%
\begin{equation}\label{sigrec}
\sigma_{nn} = \frac{2d\theta}{\eta} \frac{1}{2b_n - \alpha_n - \alpha'_n} ,
\end{equation}
where $b_n$ is given by Eq. (\ref{bn}) and the functions $\alpha_n$ and $\alpha'_n$ are found from the forward and backward recurrence relations 
%
\begin{equation}\label{alrec}
\alpha_{n+1} = \frac{1}{2b_n - \alpha_n}, \qquad
\alpha'_{n-1} = \frac{1}{2b_n - \alpha'_n}
\end{equation}
with boundary conditions $\alpha_1=\alpha'_N=0$.
The nondiagonal CFs can then be written as
%
\begin{equation}\label{signn'}
\renewcommand{\arraystretch}{1.5}
\sigma_{nn'} = \sigma_{nn} \times
\left\{
\begin{array}{ll}
\prod_{l=n}^{n'-1} \alpha'_l,                      & n'>n                          \\
\prod_{l=n}^{n'+1} \alpha_l,                      & n'<n .		\\
\end{array}
\right. 
\end{equation}
In the bulk the quantity $b$ of Eq. (\ref{bn}) is independent of $n$, and 
 Eqs. (\ref{alrec}) can be solved to give
%
\begin{equation}\label{albulk}
 \alpha = b-\sqrt{b^2-1} = \alpha' ,
\end{equation}
The solution the spin CFs above thus simplifies to the result
%
\begin{equation}\label{sigbulk}
\sigma_{nn'}^{\rm bulk}({\bf q}) = 
\frac{d\theta}{\eta} \frac{\alpha^{|n-n'|}}{\sqrt{b^2-1}},
\end{equation}
that can also be obtained with other methods.
For the numerical solution in the film geometry, diagonal CFs 
$\sigma_{nn}$ of  Eqs. (\ref{sigrec}) should be substituted into the contraint
equations (\ref{sconstrfd}), which defines the variation of $G_n$ after
application of some iterative scheme (e.g., Newton method).

There is another routine for solving system of equations (\ref{cffd}) and 
(\ref{sconstrfd}), which avoids numerical calculation of the integrals over
the Brillouin zone at each step. \cite{mih76}
The solution Eq. (\ref{cffd}) for diagonal CFs $\sigma_{nn}$ can be written in
the form
%
\begin{equation}\label{signnmat}
\sigma_{nn} = 
\frac{2d\theta}{\eta} A_{nn}^{-1} ,
\end{equation}
where $\hat A^{-1}$ is the inverse of the tridiagonal matrix $\hat A$ of
Eq. (\ref{cffd}): $A_{nn} = 2b_n$ and $A_{n,n\pm 1} = -1$.
It is convenient to represent matrix  $\hat A$ in the form
%
\begin{equation}\label{ABmat}
\hat A = \hat B -2d'\lambda'_{\bf q} \hat I ,
\end{equation}
where $\hat I$ is a unit matrix and $\hat B$ is a tridiagonal matrix:
%
\begin{equation}\label{Bmat}
B_{nn} = 2d/(\eta G_n), \qquad B_{n,n\pm 1} = -1 .
\end{equation}
Then Eq. (\ref{signnmat}) can be transformed to
%
\begin{equation}\label{signnmat1}
\sigma_{nn} = 
\frac{2d\theta}{\eta} \sum_{\rho=1}^N \frac{ U_{n\rho}^2 }
{ \lambda_\rho - 2d'\lambda'_{\bf q} },
\end{equation}
where $\hat U$ is the real unitary matrix 
($\hat U^{-1} = \hat U^T$, i.e., $U_{\rho n}^{-1} = U_{n\rho}$), that
diagonalizes $\hat B$, and $\lambda_\rho$ are the eigenvalues of $\hat B$.
The nice feature of this expression is that  $\hat U$ and $\lambda_\rho$ are
independent of ${\bf q}$.
Thus the integral over the Brillouin zone of Eq. (\ref{signnmat1}) can be
expressed through the lattice Green function of the layers, $P_{d'}$, and 
constraint equation (\ref{sconstrfd}) takes the form
%
\begin{equation}\label{sconstrfd1}
 s_{nn} = \frac{2d\theta}{\eta} \sum_{\rho=1}^N 
\frac{ U_{n\rho}^2 }{\lambda_\rho } P_{d'}(2d'/\lambda_\rho) 
= 1 . 
\end{equation}
Since $P_{d'}(X)$ can be calculated analytically or tabulated, this method
should be faster than the continued-fraction formalism described above,
especially for hypercubic lattices in high dimensions.
However, application of the latter to the semi-infinite ferromagnet in
Ref. \onlinecite{gar98pre} has shown that it is fast enough, the main problem
being the stability of iterations for low dimensionalities $d$.
For this reason, here the continued-fraction formalism is chosen for
numerical calculations, whereas the diagonalization method will be applied to
find analytical solutions for films consisting of small number of layers.   

Before proceeding to the films with free boundary conditions, let us consider
the simplified problem with periodic boundary conditions.
Here, in an artificial way, all quantities are rendered uniform, but,
nevertheless, there is a dimensional crossover discussed above.

\section{Periodic boundary conditions}
\label{pbc}

For the problem with pbc, one has $G_n=G$ and $b_n=b$ in Eq. (\ref{cffd}).
Solving the latter with boundary conditions of Eq. (\ref{bcond}) yields
%
\begin{equation}\label{sigpbc}
\sigma_{nn}^{\rm pbc}({\bf q}) = 
\frac{ d\theta }{ \eta \sqrt{b^2-1} } \frac{ 1 + \alpha^N }{ 1 - \alpha^N },
\end{equation}
where $\alpha<1$ is defined by Eq. (\ref{albulk}).
In the long-wavelength region near criticality, one has
%
\begin{equation}\label{balcrit}
\sqrt{b^2-1} \cong \sqrt{\kappa^2+q^2}, \qquad \alpha \cong 1 - \sqrt{\kappa^2+q^2} .
\end{equation}
Thus, if $N\sqrt{\kappa^2+q^2} \lsim 1$, an additional factor
$\sqrt{\kappa^2+q^2}$ appears in the denominator of  Eq. (\ref{sigpbc}), which
is responsible for the dimensional crossover in films.

The Curie temperature of a film can be found from the condition that
the longitudinal susceptibility diverges.
This amounts to vanishing of the determinant of the system of equations
for the longitudinal CF at zero wave vector, i.e., Eq. (\ref{cffd}) with $\eta=1$ and ${\bf q}=0$.
It can be seen that the latter is satisfied for $b(\eta=1, {\bf q}=0)=1$ and hence $G=1$, as in
the bulk.
Now $\theta_c$ can be found from the constraint relation (\ref{sconstrfd}).
Separating the bulk term, one can write
%
\begin{equation}\label{Tcpbc}
\theta_c^{-1} = P_d(\eta) + \frac{d'}{\Lambda^{d'}}\int_0^\Lambda dq q^{d'-1}
\frac{ 2d }{ \eta \sqrt{b^2-1} } \frac{ \alpha^N }{ 1 - \alpha^N },
\end{equation}
where, for continuous dimensionalities $d$, the lattice Green function $P_d(\eta)$
is determined by Eqs. (\ref{px}) and (\ref{contint}). 
For the hypercubic lattice, a usual integration over the Brillouin zone is
performed in Eq. (\ref{Tcpbc}).
The large-$N$ asymptotes of Eq. (\ref{Tcpbc}) can be calculated analytically.
For the weakly anisotropic model in the range $d>3$ and $N\kappa_c \ll 1$, one obtains
%
\begin{equation}\label{Tcpbc1}
\theta_c^{-1} \cong P_d(\eta) + \frac{d'}{\Lambda^{d'}}
\frac{ 2d \Gamma(d-2) \zeta(d-2) }{ N^{d-2} }.
\end{equation}
For the hypercubic lattice, one should make the replacement 
%
\begin{equation}\label{contint1}
d'/\Lambda^{d'} \Rightarrow S_{d'}/(2\pi)^{d'},
\end{equation}
where $S_d = 2\pi^{d/2}/\Gamma(d/2)$ is the surface of the $d$-dimensional unit sphere.
Dependence of such a type, derived field-theoretically, can be found in Ref. \onlinecite{o'cstebra97}.
For $3<d<4$, it is in accord with finite-size-scaling formula (\ref{TcShNbig})
with $\lambda=1/\nu_b = d-2$.
In higher dimensions for the $D=\infty$ model one has $1/\nu_b =2$, whereas
the shift exponent $\lambda$ is given by he same formula.  
A particular case of Eq. (\ref{Tcpbc1}) is \cite{barfis73,gar96jpal}
%
\begin{equation}\label{Tcpbcd4}
\theta_c^{-1} \cong P_4(1) + 2/(3N^2)
\end{equation}
for $d=4$.
For $\kappa_c N \gsim 1$, the second term in the rhs of Eq. (\ref{Tcpbc}), and
thus the shift of $T_c$, decays exponentially with $\kappa_c N$.

In three dimensions, the integral in Eq. (\ref{Tcpbc}) can be done
analytically for all values of $\kappa_c N$, provided $N\gg 1$.
For $d=3$ the result has the form\cite{gar96jpal}
%
\begin{equation}\label{Tcpbcd3}
\theta_c^{-1} \cong P_3(\eta) + \frac{3}{\pi N} 
\ln\frac{1}{1-\exp(-\kappa_cN)}.
\end{equation}
For $\kappa_c N \ll 1$ it simplifies to
%
\begin{equation}\label{Tcpbcd3lo}
\theta_c^{-1} \cong P_3(\eta) + \frac{3}{\pi N} 
\ln\frac{1}{\kappa_c N}.
\end{equation}
For the continuous-dimension lattice $d=3.0$, the second terrms of the
above equations should, according to Eq. (\ref{contint1}), be multiplied by
$\Gamma[(d+1)/2][2\pi/(d+1)]^{(d-1)/2} \Rightarrow \pi/2$.
The form of Eq. (\ref{Tcpbcd3lo})  is in accord with finite-size-scaling formula (\ref{TcShNbig}), but now
the coefficient $A$ in Eq. (\ref{TcShNbig}) depends on anisotropy in a crucial
way.
In the isotropic limit, $\kappa_c$ defined by Eq. (\ref{kapcdef}) goes to
zero, and the logarithmically divergent second term of this formula becomes
dominant.
In this limit $\theta_c$ becomes logarithmically small, in accordance with 
general formula (\ref{TcAnlo}), which for the ASM takes the form
%
\begin{equation}\label{TcAnloASM}
\theta_c^{-1}(d,\eta,N) \cong \frac{ d }{ (d-1)N } P_{d-1}\left(\frac{ d\eta-1 }{ d-1 }\right) .
\end{equation}
Indeed, the Curie temperature of the square lattice is given by
%
\begin{equation}\label{Tcd2}
\theta_c^{-1} = P_2(\eta) = \frac 1\pi \ln \frac{ 8 }{ 1-\eta } \cong 
\frac 2\pi \ln \frac{ 4\sqrt{2} }{ \kappa_c }.
\end{equation}
The extra factor $3/(2N)$ in Eq. (\ref{Tcpbcd3lo}), relative to this result,
is due to the rescaling of the exchange interaction and the number of nearest
neighbors in the mean-field transition temperature of Eq. (\ref{thetadef}).
Appearance of $N$ under the logarithm in Eq. (\ref{Tcpbcd3lo}) is due to
cutting of the integral over $q$ in Eq. (\ref{Tcpbc}) at $q\sim 1/N$ rather
than at the Brillouin-zone boundary: Only in this range of $q$ the
film behaves two dimensionally.

For $d<3$ the integral in Eq. (\ref{Tcpbc}) for $\kappa_c N \ll 1$ is dominated
by $\kappa_c \sim q \ll 1/N$, and the result has the form
%
\begin{equation}\label{Tcdlo}
\theta_c^{-1} \cong  P_d(\eta) + \frac{ d' }{ \Lambda^{d'} }
\frac{ d\kappa_c^{d-2} }{ \kappa_c N } \frac{ \pi }{ \sin[(3-d)\pi/2] } .
\end{equation}
Note that because of the small factor $\kappa_c^{d-2}$ in the numerator, both
terms of Eq. (\ref{Tcdlo}) can be comparable with each other.
For the square-lattice stripe $d=2$, the bulk term of Eq. (\ref{Tcdlo}) can be
neglected and the result simplifies to $\theta_c^{-1} = 2/(\kappa_c N)$.
This is in absolute accord with Eq. (\ref{TcAnloASM}), since for the chain 
$P_1(\eta) = 1/\sqrt{1-\eta^2}$.
For the continuous-dimension model $d=2.0$, the the expression for
$\theta_c^{-1}$ should be multiplied by $\pi/\sqrt{6} \approx 1.283$. 
It should be noted that the phase transition in chains or stripes can only
occur in the limit $D\to\infty$.
For any finite $D$, the long-range order is broken by longitudinal
fluctuations which lead to  formation of domains.

\section{Free boundary conditions}
\label{fbc}

In a film with fbc, there is an additional disordering due to boundaries that
is responsible for the larger suppression of the Curie temperature of a film
in comparison to the model with pbc.
The greatest difference between the two models arises within the MFA, where for
the pbc model $T_c$ is the same as in the bulk and for the hypercubic fbc model one has
\cite{woldewhalpal71}
%
\begin{equation}\label{TcMFA}
\theta_c = 1 - \frac 1 d \left( 1 - \cos{\frac{\pi}{N+1}} \right) .
\end{equation}
This result also can be obtained for the ASM in  the limit $\eta\to 0$,
where Eq. (\ref{cffd}) trivializes and the mean-field approximation becomes exact., 
For the pbc model in this limit, finite-size corrections to $T_c$ disappear as
$\exp(-\kappa_c N)$.

For $\eta \neq 0$, the situation for the fbc model becomes complicated, since
the gap parameter $G_n$ is nonuniform near the boundaries and cannot be
calculated analytically in most cases. 
This problem becomes, however, nonessential in high dimensions, $d>4$, as well
as in all dimensions for $\kappa_c N \gg 1$.
Here the inhomogeneity of $G_n$ is localized to the narrow region near the
film boundaries, $n \sim 1$ in the first case \cite{gar98pre} and $n\sim 1/\kappa_c \ll N$ in
the second case.
In the main part of the film, $G_n=G$ is uniform, and its value at the
critical point of the  film can be obtained from the condition that the {\em  longitudinal}
susceptibility diverges, i.e., the determinant of
system of equations (\ref{cffd}) for $\eta=1$ and ${\bf q}=0$ turns to zero.
It is instructive to calculate, instead of the determinant, the whole spin CF
from Eq. (\ref{cffd}) with the free boundary conditions of Eq. (\ref{bcond})
for the uniform $G$, ignoring the narrow boundary region where $G$ is
nonuniform.
The result for the {\em transverse} CF reads
%
\begin{equation}\label{sigfbc}
\sigma_{nn}^{\rm fbc}({\bf q}) = \frac{ d\theta }{ \eta \sqrt{b^2-1} } 
\frac{ [1 - \alpha^{2n}] [1 - \alpha^{2(N+1-n)}] }{ 1 - \alpha^{2(N+1)} },
\end{equation}
the {\em  longitudinal} CF being given by the same expression with $\eta=1$.
For the bulk CF of Eq. (\ref{sigbulk}), as well as for the CF of the pbc
model, Eq. (\ref{sigpbc}), the divergence corresponds to $b=1$.
This leads, according to Eq. (\ref{bn}) with $\eta=1$ and ${\bf q}=0$, to
$G=1$ at the phase transition point.
The pole at $b=1$ is cancelled, however, in  Eq. (\ref{sigfbc}), since,
according to  Eq. (\ref{albulk}),  $b=1$ entails $\alpha=1$.
The new pole in the fbc correlation function corresponds to $\alpha^{N+1}=-1$,
i.e., to 
%
\begin{equation}\label{albcrit}
\alpha = \exp \frac{ i\pi }{ N+1 }, \qquad b = \frac 12 (\alpha + \alpha^{-1})
= \cos \frac{ \pi }{ N+1 } .
\end{equation}
Now the value of $G$ at the Curie temperature of the film can be found from Eq. (\ref{bn}) 
with $\eta=1$ and ${\bf q}=0$.
The result has the form
%
\begin{equation}\label{Gfbccrit}
G^{-1} =  1 - \frac 1 d \left( 1 - \cos{\frac{\pi}{N+1}} \right) .
\end{equation}
Now $\theta_c$ can be found from the constraint relation (\ref{sconstrfd}) for
the {\em transverse} CF of Eq. (\ref{sigfbc}).
The integral over the Brillouin zone in Eq. (\ref{sconstrfd})  simplifies in
both cases, $d>4$ and $\kappa_c N \gg 1$.
In high dimensions, the integral is dominated by $q\sim 1$, and at such $q$
 the terms with $\alpha$ to high powers  beyond the
boundary regions can be neglected.
The same situation takes place for $\kappa_c N \gg 1$ at all $q$; in both cases the bulk
CF enters under the integral and the the Curie temperature of the film is
given by
%
\begin{equation}\label{Tcfbcgen}
\theta_c^{-1} =  GP(\eta G) ,
\end{equation}
with $G$ defined by Eq.  (\ref{Gfbccrit}).
In the Ising limit $\eta\to 0$, the mean-field result of Eq. (\ref{TcMFA})
is recovered, since $P(0)=1$. 
This result is valid for all $N$, since in this limit $G_n=G$ is uniform
throughout the film and all calculations above are exact. 
If $\eta$ substantially deviates from zero, this approach works only for large
$N$, where Eq.  (\ref{Gfbccrit}) simplifies to
%
\begin{equation}\label{Gfbccrit1}
G \cong 1 + \frac{ 1 }{ 2d } \left( \frac \pi N \right)^2 .
\end{equation}
Since the last term in this expression is a small perturbation, Eq.
(\ref{Tcfbcgen}) can be written in the form \cite{gar96jpal}
%
\begin{equation}\label{Tcquasibulk}
\theta_c \cong \frac{1}{P(\eta)}
\left[
1 - \frac{1}{2d}
\left(\frac{\pi}{N}\right)^2 I(\eta)
\right] ,
\end{equation}
where
%
\begin{equation}\label{Ieta}
I(\eta) \equiv 1 + \eta P'(\eta)/P(\eta) .
\end{equation}
Function $I_d(\eta)$ shows the same qualitative behavior as  $P_{d-2}(\eta)$
(see, e.g., Ref. \onlinecite{gar98pre}).
In particular, in the weakly anisotropic case $\kappa_c\ll 1$ [see  Eq.  (\ref{kapcdef})], one has
%
\begin{equation}\label{Ilimsd}
\renewcommand{\arraystretch}{1.2}
I(\eta) \cong
\left\{
\begin{array}{ll}
\tilde C_d/ \kappa_c^{4-d},		  & 1 \leq d \leq 4	\\
I(1) - \tilde C_d\kappa_c^{d-4},                      & 4< d<6 .
\end{array}
\right. 
\end{equation}
For $d>4$, $I(\eta)$ remains finite in the isotropic limit, $\eta\to 1$.
In this dimensionality range formula (\ref{Tcquasibulk}) works for all
anisotropies, as was argues above.
In the marginal case $d=4$,  $I(\eta)$ diverges logarithmically for $\eta\to 1$.
For the hypercubic lattice,  Eq.  (\ref{Tcquasibulk}) takes the form
%
\begin{equation}\label{Tcquasibulkd4}
\theta_c^{-1} \cong P_4(1) + \frac{1}{N^2}\ln\frac{c}{\kappa_c} , 
 \qquad c \sim 1 .
\end{equation}
If anisotropy becomes so small that $\kappa_c N \lsim 1$, then the integral
over the Brillouin zone in the constraint relation is cut at the lower limit
$q \sim 1/N$ instead of $q \sim \kappa_c$.
This leads to the result 
%
\begin{equation}\label{Tcquasibulkd4iso}
\theta_c^{-1} \cong P_4(1) + \frac{ \ln N }{ N^2 } + \frac{ C }{ N^2 },
\end{equation}
where $C\sim 1$ cannot be calculated analytically.
For the continuous-dimension lattice $d=4.0$, one should, according to
Eq. (\ref{contint1}),  insert the factor  $3\pi/(5\sqrt{10})\approx 1.873$ in
front of the logarithms in the above equations.

For $d<4$, $I(\eta)$ and thus the finite-size correction to $T_c$ given by Eq. (\ref{Tcquasibulk})
diverges in the isotropic limit, $\eta \to 1$.
In the dimensionality range $3<d<4$, however, the latter does not imply that
$T_c$ goes to zero, since the film is a system of dimension $d'=d-1$ and it
can still order without anisotropy for $d'>2$. 
In fact, the situation here is determined by small wave vectors, 
$q \lsim 1/N \ll 1$, in contrast to $q \sim 1$ for $d>4$.
In this wave-vector range $d'$-dimensionality of the film manifests itself by
the appearance of an additional power of $q$ in the denominator of the spin CF
$\sigma_{nn}({\bf q})$.
Dimensional arguments show that this should be the combination 
 $N\sqrt{\kappa^2+q^2} \lsim 1$.
Thus the finite-size correction to $T_c$ has the form
%
\begin{equation}\label{Tcd34}
\theta_c^{-1} - P_d(\eta) \; \sim \; \frac 1 N \int_0^{1/N}  
\frac{dq q^{d-2} }{ \kappa^2 + q^2 },
\end{equation}
which leads to the same dependence  $\Delta T_c \equiv T_c(\infty) -T_c(N) \propto 1/N^{d-2}$
 as for the model with pbc, Eq. (\ref{Tcpbc1}). 
Analytical calculation of the numerical factor in this dependence seems
to be very difficult or impossible, because it requires knowing of the
self-consistently determined variation of $G_n$ in the film.
This numerical factor should be larger that for the model with pbc because of
the additional disordering at the film boundaries.
One can see that for $d<4$ this contribution to the shift of $T_c$ is larger
than the contribution  $\Delta T_c  \propto 1/N^2$ from the region $q\sim 1$,
which is described by Eq. (\ref{Tcquasibulk}).
At large distances $N \gsim 1/\kappa_c$, the wave-vector range in which the
film behaves $d'$-dimensionally disappears, as can be seen from Eq. (\ref{Tcd34}). 
Here a crossover to the regime of Eq. (\ref{Tcquasibulk}) occurs.

For $d\leq 3$, the intergal in  Eq. (\ref{Tcd34}) diverges at the lower limit
in the isotropic case.
In this range, the dimensionality of the film is $d'\leq 2$ and it cannot order
without anisotropy.
For very small anisotropies, $T_c$ of the film is very low, and at such
temperatures the chains of spins going from one boundary to the other are
strongly correlated and behave as single composite spins, as was explained in the
Introduction.
Here the situation simplifies, and the type of boundary conditions no longer
plays a role in the determination of $T_c$ that is given by  Eq. (\ref{TcAnloASM}).
In the marginal case $d=3$, the corresponding formula has only logarithmic
accuracy, \cite{gar96jpal} and the more accurate result for the simple cubic
lattice has the form
%
\begin{equation}\label{Tcfbcd3lo}
\theta_c^{-1} \cong P_3(\eta) + \frac{3}{\pi N} 
\ln\frac{1}{\kappa_c N} + \frac C N ,
\end{equation}
where $C\sim 1$ should be calculated numerically [cf. Eq. (\ref{Tcfbcd3lo})] .
It is interesting to note that the variation of $G_n$ throughout the film can
be easily found in the limit under consideration.
It adjusts in a way that is compatible with the full correlation of the
film's layers, i.e.,  with $\sigma_{nn'}$ being the same for all $n$ and $n'$ at small
$q$.
The latter requires that all $\alpha_n$ and $\alpha'_n$ in Eq. (\ref{signn'})
are close to 1.
Then from  Eq. (\ref{alrec}) one finds that $b_n \cong 1$ inside the film.
On the contrary, in the boundary layers one obtains $b_1=b_N \cong \onehalf$.  
The resulting variation of $G_n$ in the film can now be determined from Eq. (\ref{bn})
and has the form
%
\begin{equation}\label{gnt0}
\renewcommand{\arraystretch}{1.5}
G_n \cong
\left\{
\begin{array}{ll}
\displaystyle
\frac{2d}{2d-1},                      & n=1,N                           \\
\displaystyle
1,		               & n\neq 1,N .
\end{array}
\right. 
\end{equation}

To conclude this section, let us consider the bilayer, $N=2$.
For the bilayer, Curie temperature can be calculated analytically with the
help of the diagonalization formalism described at the end of Sec. \ref{basic}.  
At $T_c$ one has $G_1=G_2 = 2d/(2d-1)$, and the Curie temperature ifself is
given by
%
\begin{equation}\label{Tcbilay}
\theta_c^{-1} = \sum_\pm  \frac{ d }{ 2d-1 \pm \eta } 
P_{d-1}\left(\frac{ 2(d-1)\eta }{ 2d-1 \pm \eta }\right) ,
\end{equation}
where $\sum_\pm $ sums terms with both signs.
In dimensions $d\leq 3$, the term with the sign minus diverges in the
isotropic limit, $\eta\to 1$, whereas the term with the sign plus remains
finite.
In this case the result is compatible with the general-$N$ formula
(\ref{TcAnloASM}).

For films consisting of larger number of layers, the diagonalization formalism becomes
cumbersome. 
For $N=3$, in particular, the cubic characteristic equation for the problem
simplifies, and one can write down analytical expressions for the correlation
functions in terms of $G_1=G_3$ and $G_2$.
The latter and the Curie temperature, however,  are defined by a system of
transcedental equations that cannot be solved analytically.

\section{Numerical results}
\label{numres}

The system of equations describing the anisotropic spherical model was 
solved numerically in the following way.
For a given variation of $G_n$, the transverse layer-layer autocorrelation
functions $\sigma_{nn}$ were found with the help of Eqs. (\ref{sigrec}) and (\ref{alrec}). 
The results were substituted into the constraint relations (\ref{sconstrfd})
to obtain, after the integration over ${\bf q}$, the system of nonlinear
equations for the set of $G_n$.
The latter was solved by the Newton method.
In fact, all equations were rewritten in terms of deviations from the bulk
values (see Ref. \onlinecite{gar98pre}), which helps to improve the accuracy.
This calculation yield the variation of $G_n$ throughout the film for any
temperature $T \geq T_c$.
To find the value of $T_c$, the nonlinear equation in temperature, ${\cal D}(T)=0$, 
was solved, where ${\cal D}(T)$ is the determinant of system of linear
equations (\ref{cffd})  for $\eta=1$ and ${\bf q}=0$.
In this system of equations, the values of $b_n$ are given by Eq. (\ref{bn}),
where $G_n$ are functions of temperature found on the previous step of the
numerical routine.

\begin{figure}[t]
\unitlength1cm
\begin{picture}(11,6.5)
\centerline{\epsfig{file=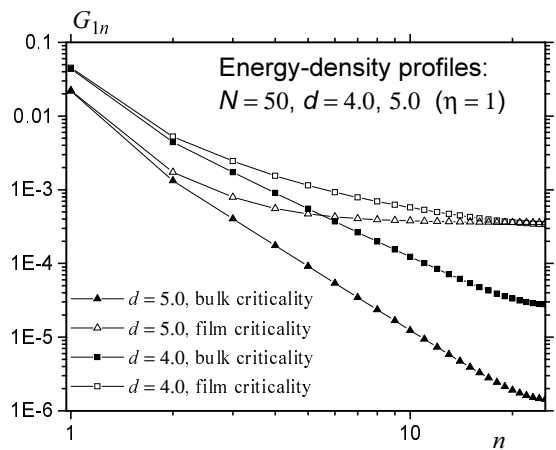,angle=-90,width=8cm}}
\end{picture}
%
\caption{ \label{sff_gd45}
Energy-density profiles at bulk and film criticalites for the isotropic
$D=\infty$ model in four and five dimensions.
}
\end{figure}
\begin{figure}[t]
\unitlength1cm
\begin{picture}(11,6.5)
\centerline{\epsfig{file=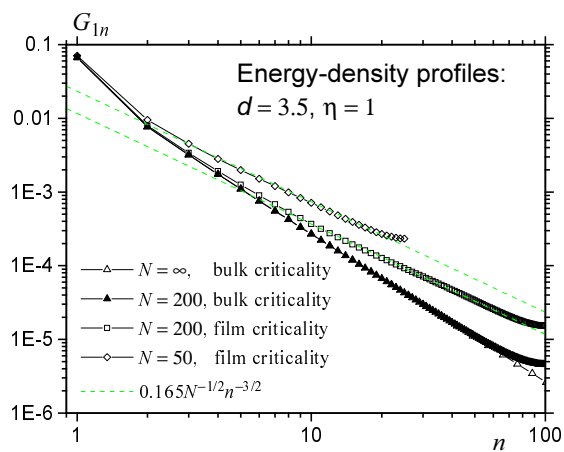,angle=-90,width=8cm}}
\end{picture}
%
\caption{ \label{sff_gd35}
Energy-density profiles at bulk and film criticalites for the isotropic
$D=\infty$ model in dimension $d=3.5$.
}
\end{figure}
\begin{figure}[t]
\unitlength1cm
\begin{picture}(11,6.5)
\centerline{\epsfig{file=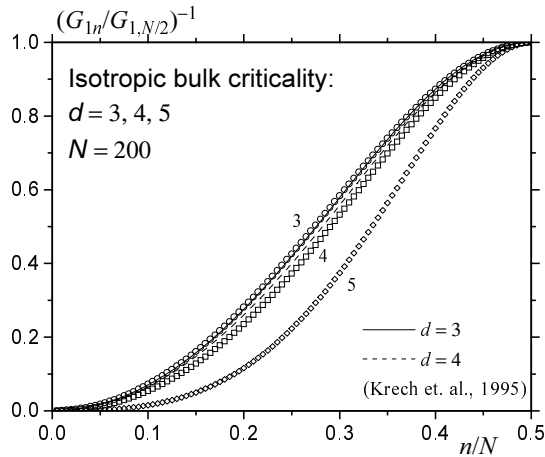,angle=-90,width=8cm}}
\end{picture}
%
\caption{ \label{sff_g345}
Reciprocal of normalized energy-density profiles at bulk criticalites for the isotropic
$D=\infty$ model in dimensions $d=3,4$ and 5.
}
\end{figure}

It is convenient to plot, instead of $G_n$, its deviation from the bulk value 
$G_{1n} \equiv G_n - G$ which at and below the bulk criticality (i.e., for
$G=1$) is equal to the nonuniform part of the reduced energy
density $\tilde U_{1n}$. \cite{gar98pre}
The variations of $G_{1n}$ in the isotropic films at bulk and film
criticalities in four and five dimensions for the continuous-dimension model
are shown in Fig. \ref{sff_gd45}.
The results for the hypercubic lattice differ from the latter by nonuniversal
factors of order unity and are not shown.
One can see that for $d=5$ at the film criticality the value of $G_{1n}$ is
close to $(\pi/N)^2/(2d)\approx 4 \times 10^{-4}$ in the main part on the film, 
in accordance with Eq. (\ref{Gfbccrit}).
The bulk-criticality profiles of $G_{1n}$ that are plotted for comparison,
show a very different behavior.
They follow the appropriate semi-infinite profiles $G_{1n} \sim 1/n^{d-2}$
(Ref. \onlinecite{gar98pre}) until approximately $n \sim N/4$.
Crossing of the latter with the plato level at some $n^* \sim N^{2/(d-2)}$ determines the
boundary region $n \lsim n^*$ where the semi-infinite $G_{1n}$ profile is realized.  
For $d>4$, the value of $n^*$ increases slowlier that $N$; thus for thick films
the plato of $G_{1n}$ dominates the film's Curie temperature, as was explained
in Sec. \ref{fbc}
In the marginal case $d=4$, the plato of $G_{1n}$ is much less pronounced.
In fact, it arises only for $\ln N \gg 1$.

For $d<4$, the problem becomes more complicated, and no analytical solution
for the variation of $G_{1n}$ could be obtained.
Numerical results for $d=3.5$ that are shown in Fig. \ref{sff_gd35} are also different for film and
bulk criticalities.
At the film criticality, the power-law dependence 
$G_{1n} \approx 0.165 N^{-1/2} n^{-3/2}$ is seen, instead of a plato.
For comparison, the dependence of $G_{1n}$ in the semi-infinite geometry,
which is given by Eq. (\ref{g1nd24}), has the form $G_{1n} \approx 0.0268  n^{-2}$
for $d=3.5$.
It is clear that for very thick films both criticalities are very close to
each other, and, at the film criticality, the semi-infinite solution should be realized in some region
not too far from the boundary.
The crossover from this solution to the numerically found $n^{-3/2}$  law can
be located by equating the  expressions above.
The result is that for $d=3.5$ at the film criticality, the semi-infinite
solution is realized in the region $1 \ll n \lsim n^*$ with $n^* \approx 0.0264 N$, 
whereas the $n^{-3/2}$
law holds for $n^* \lsim n$ but not too close to the center of the film
(see Fig. \ref{sff_gd35}).
What is the nature of the numerical parameter 0.0264?
Is it small enough to allow an analytical derivation of the $n^{-3/2}$ law? 
These questions have not been addressed in this article.

For $d\leq 3$, the film's Curie temperature and the energy-density profile at
the film criticality depend on anisotropy in an essential way.
This makes the corresponding figures less interesting; in particular, for $\kappa_c N \gg 1$
there is a plato of $G_{1n}$ at the distances from the boundaries $n \gsim 1/\kappa_c$.
By contrast, energy-density profiles at the bulk criticality can be
studied for $d>2$ for isotropic models.
This problem for the semi-infinite and film geometries can be addressed with
field-theoretical methods for systems with an arbitrary number of spin
components $D$ (see, e.g., Refs. \onlinecite{die86a} and \onlinecite{die97}).
In Refs.  \onlinecite{eiskredie93} and \onlinecite{kreeisdie95}, the energy-density
profiles in critical films with various boundary conditions have been
calculated with the help of the $\epsilon$ expansion.
For the free boundary conditions considered here,  the
exponentiated energy-density profile $g$ is given by Eq. (5.7) of  Ref.  \onlinecite{kreeisdie95}.
In the limit $D\to\infty$ one can use $\nu=1/2$ and $\nu=1$ in four and three
dimensions, respectively, and for $d=4$ and $d=3$ the result of Ref.  \onlinecite{kreeisdie95}
simplifies to
%
\begin{eqnarray}\label{eprofeps}
&&
g_4(x) =  \frac{ \pi^2 }{ \sin^2\pi x } -  \frac{ \pi^2 }{ 3 }
\nonumber\\
&&
g_3(x) = -[\psi(x) + \psi(1-x) +2\gamma ]  \frac{ \pi }{ \sin\pi x } -  \frac{ \pi^2 }{ 6 },
\end{eqnarray}
where $z$ is the distance from the left boundary, $L$ is the film thickness,
$x\equiv z/L=n/N$, $\psi(x) \equiv \Gamma'(x)/\Gamma(x)$, and $\gamma=0.57721566\ldots$
The quantity $g(x)$, by definition, may differ from $G_{1n}$ only by a numerical
factor which is dependent on $N$.
Thus the reduced quantity $\tilde g(x) \equiv g(x)/g(0.5)$ can be compared with the $G_{1n}$ profiles numerically calculated for the
isotropic $D=\infty$ model, which are shown in the scaled form in Fig. \ref{sff_g345}.
One can see that for $d=3$ the agreement is reasonably good, keeping in mind
that $g_3(x)$ of Eq. (\ref{eprofeps}) has been obtained in the first order in
$\epsilon=4-d$.
The curve $1/\tilde g_4(x)$, however, goes in the middle between the  curves
representing the numerical results for $d=4$ and $d=3$. 
The origin of this discrepancy is that  $g_4(x)$ of Eq. (\ref{eprofeps})
does not comprise logarithmic terms which appear in four dimensions. 
In particular, for the semi-infinite problem in four dimensions one has \cite{gar98pre}
$G_{1n} = [16n^2 \ln(an)]^{-1}$ with $a\sim 1$ for $n\gg 1$.
By contrast, both $g_4$ and $g_3$ simplify to $g(x) =1/x^2$ for $x\ll 1$,
which corresponds to the the semi-infinite case.
For $d>4$, the variation of $G_{1n}$
can be found analytically, at least at or above the bulk criticality, by the
integration of the free proparator, as was done for the semi-infinite problem
in Sec. IIIC of Ref. \onlinecite{gar98pre}.
This routine coincides with that used in Ref. \onlinecite{kreeisdie95}
in the film geometry up to numerical factors; 
thus the results of both approaches, in the scaled form, are identical in high dimensions.
For $d=4$, however, logarithmic term appears in the method of Ref.
\onlinecite{gar98pre}, in contrast to that of Ref. \onlinecite{kreeisdie95}. 

Another possible normalization of the energy-density profiles is that with
respect to the solution of the semi-infinite problem $G_{1n}^\infty$.
The function $\bar g(n/N) \equiv G_{1n}/G_{1,N/2}^\infty$ contains more information
than $\tilde g \equiv G_{1n}/G_{1,N/2}$ used above.
In particular, the value of $\bar g$ in the middle of the film can be used to
compare the accuracy of different approaches. 
So, for $d=3$ the numerical calculation yields  $\bar g_3(0.5) \approx 1.822$.
For $d=4$, the value of $\bar g(0.5)$ slightly increases with $N$ because of
the logarithmic corrections to scaling: One obtains $\bar g_4(0.5)\approx 1.772$
for $N=100$ and $\bar g_4(0.5)\approx 1.795$ for $N=200$.
On the other hand, the results of Ref. \onlinecite{kreeisdie95} are scaled as
$\bar g(x) = g(x)/4$; thus from Eq. (\ref{eprofeps}) one obtains 
$\bar g_3(0.5)\approx 1.766$ and $\bar g_4(0.5)\approx 1.645$.
One can see that the discrepancy between the exact solution for the $D=\infty$ model presented
here and the field-theoretical results of Ref. \onlinecite{kreeisdie95} is
comparable to the difference between the results for $d=3$ and $d=4$.

It is interesting to note that the numerical solution for $d=3$ can be represented with a pretty good
accuracy by the empirical formula
%
\begin{equation}\label{G1nd3fit}
\bar g(n/N) \equiv \frac{ G_{1n} }{ G_{1,N/2}^\infty } = 
\frac{ (\pi/2)^{4/3} }{ \sin^{4/3}[(\pi/2)(2x)^{3/2}] }.
\end{equation}
In Fig. \ref{sff_g345}, the corresponding curve would go exactly over the circles
for $d=3$, so it is not shown.
The value $\bar g(0.5)= (\pi/2)^{4/3} \approx 1.826$ is very close to the
numerically calculated value 1.822 cited above. 
\begin{figure}[t]
\unitlength1cm
\begin{picture}(11,6.5)
\centerline{\epsfig{file=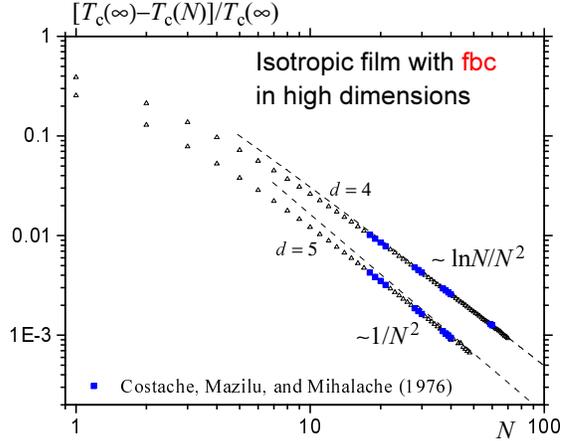,angle=-90,width=8cm}}
\end{picture}
%
\caption{ \label{sff_tc45}
Curie-temperature shift in films with fbc for the isotropic
$D=\infty$ model in four and five dimensions (continuous-dimension model).
}
\end{figure}
\begin{figure}[t]
\unitlength1cm
\begin{picture}(11,6.5)
\centerline{\epsfig{file=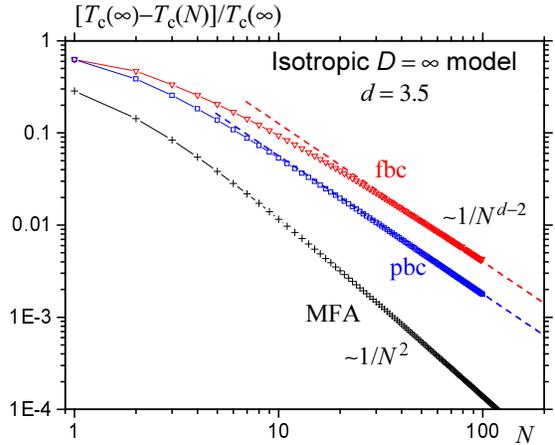,angle=-90,width=8cm}}
\end{picture}
%
\caption{ \label{sff_tc35}
Curie-temperature shift in films with fbc for the isotropic
$D=\infty$ model in dimension $d=3.5$.
}
\end{figure}
\begin{figure}[t]
\unitlength1cm
\begin{picture}(11,6.5)
\centerline{\epsfig{file=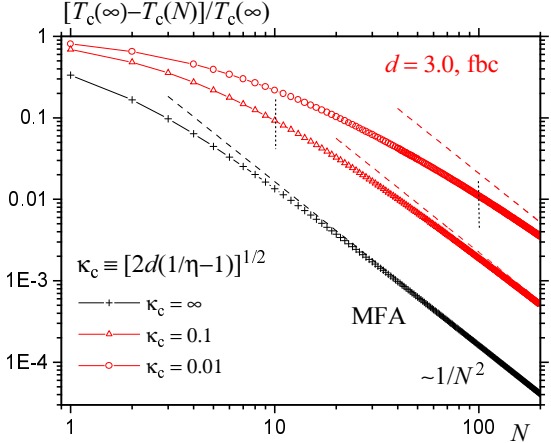,angle=-90,width=8cm}}
\end{picture}
%
\caption{ \label{sff_tc30}
Curie-temperature shift in  films with fbc for the ASM
 in three dimensions (continuous-dimension model).
The crossover values of $N$ corresponding to $\kappa_c N =1$ are shown by
short vertical lines.
}
\end{figure}
\begin{figure}[t]
\unitlength1cm
\begin{picture}(11,6.5)
\centerline{\epsfig{file=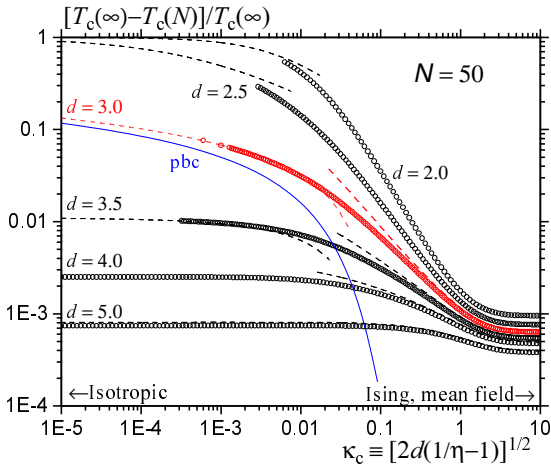,angle=-90,width=8cm}}
\end{picture}
%
\caption{ \label{sff_tckp}
Curie-temperature shift in  films with fbc for the ASM in different dimensions
vs anisotropy  (continuous-dimension model).
High- and low-anisotropy asymptotes are shown by dashed lines.
}
\end{figure}

Numerical results for the relative shift of $T_c$ in isotropic films with free
boundary conditions in high dimensions are shown in Fig. \ref{sff_tc45}.
For $d=5$, the data approach at large $N$ the asymptotic formula (\ref{Tcquasibulk}) with
$I_5(1) \approx 1.66$. 
For $d=4$, the results for thick films follow Eq. (\ref{Tcquasibulkd4iso})
with $C\approx 1.5$.
Both of these asymptotes are shown by dashed lines in Fig. \ref{sff_tc45}. 
For both hypercubic lattices, the data are in accord with those of
Ref. \onlinecite{cosmazmih76}, where a spherical model with separate spin
constraints in each of the layers was used. 
This modified spherical model is thus better than the standard one using the
global spin constraint.
However, it cannot be generalized for the anisotropic case and applied to 
films in three dimensions.

To illustrate the situation in the range $3<d<4$ where the isotropic
$D=\infty$ model shows nontrivial power law for the $T_c$ shift in films, 
the data for $d=3.5$ are shown in Fig. \ref{sff_tc35}.
One can see that both models with pbc and fbc show the same shift exponent 
$\lambda=d-2$.
For the model with pbc, the amplitude $A$ in Eq. (\ref{TcShNbig}) calculated from
Eq. (\ref{Tcpbc1}) is $A \approx 1.788$.
For the fbc model, one obtains $A\approx 4$ from the fit to the numerical
results.
This value is substantially larger than for the model with pbc because of the
boundary effects that affect the whole sample (see Fig. \ref{sff_gd35}). 
Also due to the boundary effects, the data approach the power-law asymptote
much slowlier than for the pbc model.

Numerical results for the $T_c$ shift in films with fbc in there dimensions
for different values of the anisotropy are shown in Fig. \ref{sff_tc30}.
The data follow Eq. (\ref{Tcfbcd3lo}) at $\kappa_c N \lsim 1$ and 
Eq. (\ref{Tcquasibulk}) for larger thicknesses, $\kappa_c N \gsim 1$.
In the latter case, the shift exponent in Eq. (\ref{TcShNbig})  is 
$\lambda = 2$, as in the mean-field case, but the amplitude $A$ can be very
large for small anisotropy.

Finally, the dependences of the $T_c$ shift in films of $N=50$ layers in different dimensions
are shown as function of the anisotropy in Fig. \ref{sff_tckp}.
For $d=5$, the data are well described by the formula (\ref{Tcquasibulk})
anisotropy range.
For $d=4$, this formula fails at small anisotripies because of the logarithmic
divergence of $I_4(\eta)$ at $\eta=1$ and it should be replaced by 
Eq. (\ref{Tcquasibulkd4iso}) in the isotropic limit.
For $d=3.5$, the Newton algorithm for finding $T_c$ fails at small
anisotropies because the inverse longitudinal susceptibility of a film or the
determinant of the linear system of equations (\ref{cffd}) approach zero very
flat:
$\chi_z^{-1}(T) \propto {\cal D}(T) \propto (T-T_c)^{2/(d-3)}$
in the isotropic case.
Knowing this dependence, one can go over to the equation for the nonlinearly
scaled quantity, e.g., ${\cal D}^{(d-3)/2}$, that behaves linearly near $T_c$.
In such a way, in fact, the data in Fig. \ref{sff_tc35} have been obtained.
This yields $T_c(N)/T_c(\infty) \approx 0.989$ for $d=3.5$ in the isotropic
limit.
The functional form of the dependence of $T_c$ on anisotropy near the
isotropic limit can be found from the observation that the film behaves as a
$d'$-dimensional system in the wave-vector range $qN \lsim 1$.
One can write [cf. Eq. (\ref{thetacbulk})]
%
\begin{equation}\label{Tcd35Anlo}
\frac{ T_c(N) }{ T_c(\infty) } \cong \frac{ T_c(N) }{ T_c(\infty) }\bigg|_{\kappa_c=0} 
\times \left( 1 + \frac{ c_d \kappa_c^{d-3} }{ N } \right),
\end{equation}
where $c_d$ should be determined numerically.
Fit to the data for $d=3.5$ yields $c_{3.5} \approx 2.40$.
With this choice, asymptote  (\ref{Tcd35Anlo}) overlaps with the numerical
results in Fig. \ref{sff_tckp} in a sufficiently wide range of $\kappa_c$.

In three dimensions, $T_c(N)$ goes to zero logarithmically in the isotropic
limit according to Eq. (\ref{Tcfbcd3lo}).
The fit to the numerical data for the continuous-dimension lattice [$3/\pi$ in
front of the logarithm in  Eq. (\ref{Tcfbcd3lo}) is replaced by 3/2] yields
$C=1.77$.
Again, this asymptote well fits the numerical data in a wide range of anisotropies.

For $d<3$, the Curie temperature of the film goes to zero as $T_c \propto N
\kappa_c^{3-d}$ in the isotropic limit, in accordance with Eq. (\ref{Tcdlo}).
The corresponding asymptotes for $d=2.5$ and $d=2.0$ are shown by dashed lines
in Fig. \ref{sff_tckp}.
Unfortunately, numerical calculations could not been performed for such small
anisotropies because of the instability of the numerical algorithm.

\section{Discussion}
\label{discussion}  

In this paper a profound influence of the anisotropy on ordering in ferromagnetic
films has been investigated.
The main result is that in three dimensions, the behavior of the film of
arbitrary number of monolayers $N$ can be dominated, for sufficiently small
anisotropy, by two-dimensional fluctuations which strongly suppress the
ordering temperature.
{\em Films described by the isotropic Heisenberg model cannot order in three
dimensions}.
Most of results above have been obtained for the exactly solvable
infinite-component spin-vector model, but more realistic models as the
Heisenberg model should share the same qualitative behavior.
In particular, the small-anisotropy formula for $T_c$, Eq. (\ref{TcAnlo}),
 should be valid for all models in dimensions less than three.
In this dimensionality range, the shift amplitude $A$ in Eq. (\ref{TcShNbig})
should diverge in the isotropic limit for all systems. 

The theoretical tool used here, the anisotropic spherical model, allows to
separate the major effects of Goldstone or would be Goldstone modes in systems
of reduced dimensionality from usually less significant but complicative effects
of critical fluctuation coupling.
The ASM has a lot of advantages in comparison to the mean-field approximation
or to the standard spherical model, and it should be tried for classsical spin
systems the next after the MFA.
Some more subtle effects, as the Berezinsky-Kosterlitz-Thouless phase transition, are,
however, not reproducible within the ASM.

A peculiar feature of the ASM is that only fluctuations transverse with
respect to the easy axis give contribution to the equations describing static
properties of the system.
Thus these equations contain the transverse correlation length $\xi_{c\perp}
\equiv 1/\kappa$ as the length scale.
In particular, crossover between different regimes for the Curie temperature
of the film, $T_c(N)$, occurs at $\kappa_cN \sim 1$, where $\kappa_c$ is given
by Eq. (\ref{kapcdef}).
The longitudinal correlation function and longitudinal correlation length are the ``slave''
quantities that are decoupled from the system of equations for the ferromagnet
and can be determined after the solution of the latter.
By contrast, in the conventional theory of phase transitions the diverging
longitudinal correlation length is used as the length scale.
It would be interesting to see how does it appear in the theory of ordering in
films, if one goes beyond the $D=\infty$ model, i.e., takes into account the
$1/D$ corrections.

A more urgent task, however, is to solve the system of equations describing
the anisotropic spherical model below $T_c$, both in the semi-infinite and
film geometries.

To conclude, it should be noted that equations resembling those describing the
ASM can be obtained for quantum systems in spirit of the random-phase
approximation by decoupling of the high-order Green functions.
Examples of such works in the film geometry are Refs. \onlinecite{mih76} and  
\onlinecite{irkkatkat96}.


\begin{thebibliography}{10}

\bibitem[*]{e-gar}
Permanent address: I. Institut f\"ur Theoretische Physik, Universit\"at
Hamburg, Jungius Strasse 9, D-20355 Hamburg, Germany. 

Electronic addresses:  garanin@mpipks-dresden.mpg.de\\
garanin@physnet.uni-hamburg.de\\
http://www.mpipks-dresden.mpg.de/$\sim$garanin/ \\


\bibitem{neu62}
{C. A. Neugebauer}, Phys. Rev. {\bf 116},  1441  (1962).

\bibitem{gra74}
{U. Gradmann}, Appl. Phys. {\bf 3},  161  (1974).

\bibitem{dueetal89}
{W. D\"urr, M. Taborelli, O. Paul, R. Germer, W. Gudat, D. Pescia, and M.
  Landolt}, Phys. Rev. Lett. {\bf 62},  206  (1989).

\bibitem{libab92}
{Yi Li and K. Baberschke}, Phys. Rev. Lett. {\bf 68},  1208  (1992).

\bibitem{schsierie96}
{P. Schilbe, S, Siebentritt, and K.-H. Rieder}, Phys. Lett. A {\bf 216},  20
  (1996).

\bibitem{bab97}
{K. Baberschke}, Appl. Phys. A {\bf 62},  417  (1997).

\bibitem{woldewhalpal71}
{T. Wolfram, R. E. Dewames, W. F. Hall, and P. W. Palmberg}, Surf. Sci. {\bf
  28},  45  (1971).

\bibitem{all70}
{G. A. T. Allan}, Phys. Rev. B {\bf 1},  352  (1970).

\bibitem{fis71}
{M. E. Fisher},  in {\em Critical {P}henomena}, edited by M.~S. Green (Academic
  Press, New York, 1971), Vol.~2.

\bibitem{fisbar72}
{M. E. Fisher and M. N. Barber}, Phys. Rev. Lett. {\bf 28},  1516  (1972).

\bibitem{bar83ptcp}
{M. N. Barber},  in {\em Phase Transitions and Critical Phenomena}, edited by
  C. Domb and J.~L. Lebowitz (Academic Press, New York, 1983), Vol.~8.

\bibitem{sta68pr}
{H. E. Stanley}, Phys. Rev. {\bf 176},  718  (1968).

\bibitem{berkac52}
{T. N. Berlin and M. Kac}, Phys. Rev. {\bf 86},  821  (1952).

\bibitem{barfis73}
{M. N. Barber and M. E. Fisher}, Ann. Phys. (N.Y.) {\bf 77},  1  (1973).

\bibitem{kno73}
{H. J. F. Knops}, J. Math. Phys. {\bf 14},  1918  (1973).

\bibitem{cosmazmih76}
{G. Costache, D. Mazilu, and D. Mihalache}, J. Phys. C {\bf 9},  L501  (1976).

\bibitem{garlut84d}
{D. A. Garanin and V. S. Lutovinov}, Solid State Commun. {\bf 50},  219
  (1984).

\bibitem{gar97zpb}
{D. A. Garanin}, Z. Phys. B {\bf 102},  283  (1997).

\bibitem{garcan98prb}
{D. A. Garanin and B. Canals}, Phys. Rev. B  (submitted, cond-mat/9805362).

\bibitem{gar96jpa}
{D. A. Garanin}, J. Phys. A {\bf 29},  2349  (1996).

\bibitem{koegarharjah93}
{J. K\"otzler, D. A. Garanin, M. Hartl, and L. Jahn}, Phys. Rev. Lett. {\bf
  71},  177  (1993).

\bibitem{gar98pre}
{D. A. Garanin}, Phys. Rev. E  (in press, cond-mat/9803230).

\bibitem{gar96jpal}
{D. A. Garanin}, J. Phys. A {\bf 29},  L257  (1996).

\bibitem{bramoo77prljpa}
{A. J. Bray and M. A. Moore}, Phys. Rev. Lett. {\bf 38},  735  (1977);
 J. Phys. A {\bf 10},  1927  (1977).

\bibitem{mih76}
{D. Mihalache}, Phys. Lett. A {\bf 59},  295  (1976).

\bibitem{o'cstebra97}
{D. O'Connor, C. R. Stephens, and A. J. Bray}, J. Stat. Phys. {\bf 87},  273
  (1997).

\bibitem{die86a}
H.~W. Diehl,  in {\em Phase Transitions and Critical Phenomena}, edited by C.
  Domb and J.~L. Lebowitz (Academic Press, London, 1986), Vol.~10, pp.\
  75--267.

\bibitem{die97}
H.~W. Diehl, Int.\ J.\ Mod.\ Phys.\ B {\bf 11},  3503  (1997).

\bibitem{eiskredie93}
{E. Eisenriegler, M. Krech, and S. Dietrich}, Phys. Rev. Lett. {\bf 70},  619
  (1993).

\bibitem{kreeisdie95}
{M. Krech, E. Eisenriegler, and S. Dietrich}, Phys. Rev. E {\bf 52},  1345
  (1995).

\bibitem{irkkatkat96}
{V. Yu. Irkhin, A. A. Katanin, and M. I. Katsnelson}, J. Magn. Magn. Mater. {\bf 164},  66
  (1996).

\end{thebibliography}

\end{document}